# Various Applications of Transition Metal Dichalcogenides

Jeong Hyun Jung

†Division of Materials of Science and Engineering, Hanyang University, 222 Wangsimni-ro, Seongdong-gu, Seoul 04763, Republic of Korea  jjh8996614@naver.com (Please contact here firstly)

**ABSTRACT:** TMDs have recently been spotlighted due to their original features. Firstly, they have layered structures with Van der Waals interactions. Secondly, they have different phases, affecting a degree of anisotropy. Finally, they have excellent carrier mobilities. In this review, based on these characteristics, this journal shows how TMDs can be applied in a wide variety of industries. The main applications that will be treated is biomedical and electronic applications.

## 1 INTRODUCTION

After the discovery of TMDs, they pay a fascinating attention due to their excellent mechanical, electronic, optoelectronic properties. There are several reasons for these properties. First, X-M-X layer is bonded with Van der Waals interactions. Second, TMDs have great carrier mobility, improving integration levels and decreasing short channel effect (or reducing scattering effect). Third, TMDs can change from indirect band gap to direct band gap, which facilitates optoelectronic application.

Until now, 2D TMDs are applied for drug-delivery or substitutes for MOSFET. By using large surface and great mobility characteristics, they begin to be gradually absorbed in near future society. Nowadays, a great deal of undergraduate studies about how to adapt in frequently changing society. As a student majoring in MSE, it seems to be inevitable to study TMDs, future materials.

In order to help understand, this journal will briefly mention rudimentary characteristics of TMDs, and then explain emerging applications of them. Characteristics will include chemical formula, novel structures, and distorted phase. Applications will contain drug-delivery, MOSFET substitutes, and sensors.

## 2 Characteristics of TMDs

TMDs are composed of transition metal M (V, Nb, Ta) and two chalcogens X (S, Se). M layer is sandwiched by X layers, and both organize a wide variety of atom arrangements. The structures of TMDs, can be categorized as trigonal prismatic (with hexagonal symmetry H), octahedral (with tetragonal symmetry T), and distorted phase(T`). Most of TMDs have a shape of $MX_2$, except several original shapes like $M_2X_3$ and MX. In H symmetry, each M have six branches which make two tetrahedrons in +z and -z directions (Fig1). Along the z directions, TMDs show chalcogenide-transition metal-chalcogenide arrangement bonded by weak Van der Waals bond. Owing to low strength of bonds, bulk TMDs can be easily exfoliated to make a single layer, which is the onset of TMDs' properties. The single layer is made up of 3 atoms with 0.7nm height, virtually same as 2D materials. Because of TMDs' planar feature, TMDs have a great surface-to-volume ratio readily accommodating access for large area functionalization. Simply put, TMDs' large area allows broad interactions with target material, increasing sensitivity of it (Fig2).

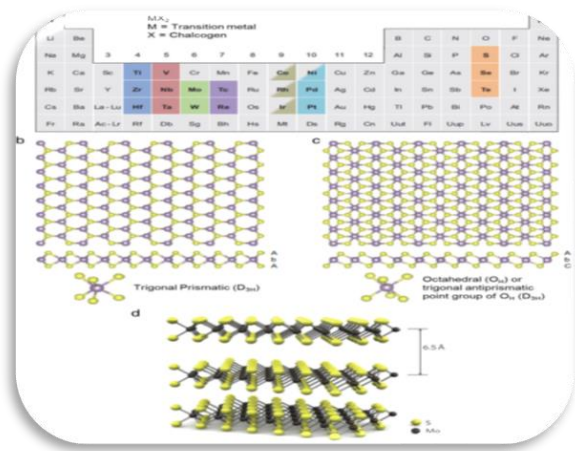

Fig1: Rudimentary information of TMDs

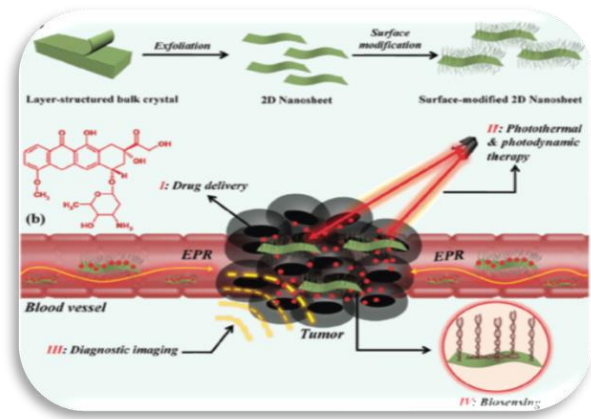

Fig2: 2D TMDs surrounding target materials

As mentioned in the first paragraph, small changes in stacking can result in different phases: 1T` phase ($ReS_2$, $ReSe_2$). This symmetry-reducing phase improves anisotropy in a conventional cell. Thus, 1T` phase shows better optoelectronic applications than symmetry phases. (THIS WILL BE TREATED LATER.) In contrast to symmetry TMDs like $MOS_2$, distorted phases ($ReS_2$) do not exhibit equal displaced chalcogenide atoms. In this case, remaining electrons in the d orbital of Re atom lead to a strong metallic Re-Re bond (Fig3). The radius of this bond is far shorter than that of Re atoms in a single crystal. This factor is a major rationale of 1T` phase's anisotropy. In addition, due to weak interlayer interactions between Re atoms, Van der Waals forces and binding energies are very weak, which cause astonishing anisotropic properties at near infrared frequencies. On the other hand, ReS$_2$ keep a constant band gap 1.58eV from bulk to monolayer. The monolayer possesses a great in-plane anisotropy with very different quantum confinement effects and a Raman spectrum, showing a proper application at optoelectronic field.

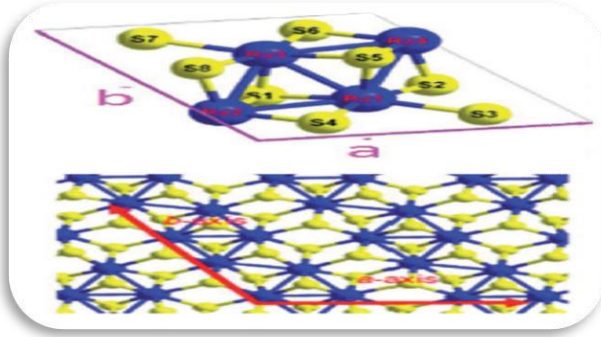

Fig3: Re-Re metallic bonds

### 3.1 Biomedical application of TMDs

TMDs are utilized for packing target material to reach a therapeutic goal. In atomic view, thin TMDs have a high photothermal response, and moderate conductivity that is proper for biomedical application. However, they are not originally bio-suitable, and require functionalization process to well act in a biological system. Functionalization can be accomplished by diagnostic tool (to increase sensitivity) or adding new abilities in TMDs like drug delivery. Generally, functionalization has two types: non-covalent and covalent. Drug delivery mechanism usually uses non-covalent functionalization. Non-covalent functionalization is used when keeping intrinsic properties of TMDs is pivotal or removal of functional group is important. This functionalization is physically absorption of drug on large TMDs' surface, maximizing loading of drug and interactions with surroundings. Covalent functionalization requires strong anchors with functional group (Fig4). By making chemical bonds of chemical groups, functional groups are easily attached to TMDs' surface, and even physical or chemical properties of TMDs can be changed. Polymers and small organic materials are proper to be covalently/non-covalently bonded on TMDs` surface for biomedical purpose.

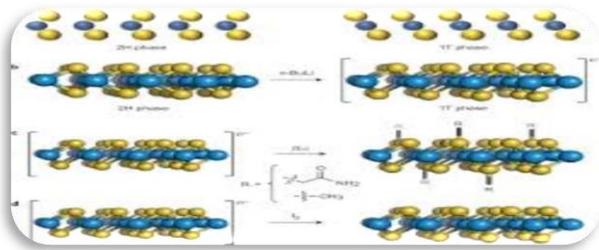

Fig4: Covalent functionalization

### 3.2 Optoelectronic & Electronic application of TMDs

Nowadays, semiconductor market pursues minimization of devices in order to construct micro-component, and MOSFETs (metal oxide semiconductor field effect transistors) are cores in microelectronic circuits. However, as the size of devices decreases, the MOSFETs face with short channel effects (which are inevitable).

Using TMDs can be a great resolution. They have dangling bond free morphology and nice atomic thickness. Hence, they have low atomic scattering effect, and mobility of carriers can be manipulated via gate voltage.

Specifically, below 10nm channel length might be challenges in conventional FETs because of drain-induced barrier lowering, surface scattering, and hot electron effects. Most MTe$_2$ based FETs have 1~68 $cm^2 \cdot V^{-1} \cdot s^{-1}$, and MOTe$_2$ (hole-doped) transistors have 20 $cm^2 \cdot V^{-1} \cdot s^{-1}$ at room temperature. In addition to mobility, MOTe$_2$ shows a $10^6$ on/off ratio. If these transistors are tuned by ferroelectric polymer poly, maximum carrier mobility is up to 68 $cm^2 \cdot V^{-1} \cdot s^{-1}$. This astound high mobility proves MOTe$_2$ transistors improve the field effect performance.

Meanwhile, 2D TMDs in photodetectors (Fig5) have two main photon-matter interactions: the photoconduction mode and the photocurrent mode based on the photovoltaic effect. In the case of the former, excited carriers directly increases device's conductance. Conversely, the latter's carriers transform into current under an asymmetric electrical field. For instance, MOS$_2$ displays 880 $A \cdot W^{-1}$ at 561nm wavelength, which is a high responsivity.

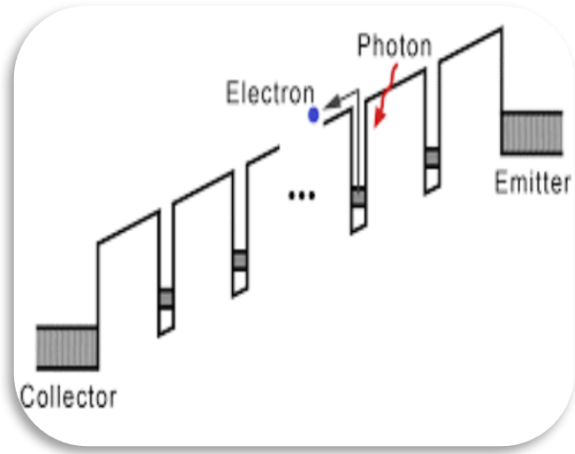

Fig5: Simplified photodetectors

Furthermore, TMDs can be applied to piezoelectricity. Piezoelectricity (Fig6) is that by accumulating polarization change from mechanical stress, stress transforms to electricity. There are several reasons why TMDs are a good candidate for piezoelectricity application. First, their non symmetry and low dimensional shape cause piezoelectricity. Second, due to their high crystallinity, they are easy to withstand stress. Third, TMDs can retain their single layer with small surface energy.

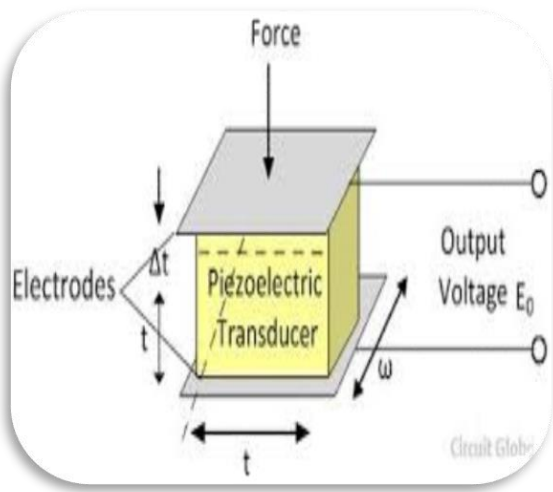

Fig6: Description of piezoelectricity

Similar with piezoelectricity, TMDs exhibit thermoelectricity properties. Thermoelectric devices convert heat into electrical energy with the Seebeck effect. The low dimension material like quantum dots have excellent thermoelectric figure of merit (ZT). For example, SnSe, showing low-dimension and anisotropic structure, has a high ZT and thermoelectric efficiency.

### 3.3 Miscellaneous applications of TMDs

TMDs' large surface area due to sheet-like structure can apply to capacitance storage and sensors. On average, TMDs have $330 F*cm^{-3}$ capacitance with $2m*W*h*cm^{-3}$, and show high sensitivity of 74 for PH. Unlike electric sensor, TMD sensors do not have physical gates to selectively absorb targeted gases. Additionally, TMDs are used as good photonic devices, especially, GaAs/h-BN solar cell (Fig7). Their power conversion efficiency is over 9%.

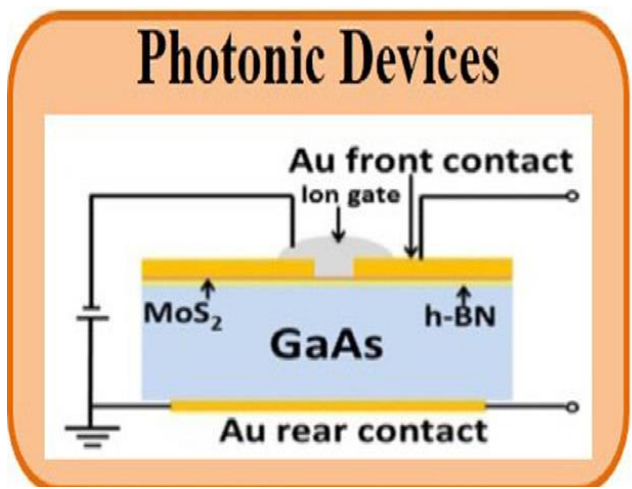

Fig7: GaAs/h-BN solar cell

## 4 Conclusion and Future outlook

This review paper outlined a few important characteristics and following applications such as biologic, optoelectronic, photonic usage. These results are derived from TMD layers' bond which is made of weak Van der Waals bond. Due to this factor, TMDs can easily be 2D, which has large surface to volume ratio. Then, the ratio helps grab target materials simply in order to deliver them.

Even though 2D TMDs are emerged recently, their development is processing rapidly, and their requirements have been skyrocketing. Therefore, in near future, they will be main products in a wide variety of material markets. For example, Samsung who is a representative of memory market with lots of prestige and finance recently tries to reduce design rule for the sake of enhancement of integration. As mentioned earlier, 2D TMDs are thin and have high carrier mobility, and their properties are suitable for materials in integrated circuits, changing paradigm of semiconductor markets. Moreover, their proper in-plane crystallinity induces piezoelectricity properties. This proves that TMDs can be a sensor not only chemically, but also physically as well. Overall, only based on anisotropic and weak bond properties, TMDs will lead a new trend of materials.

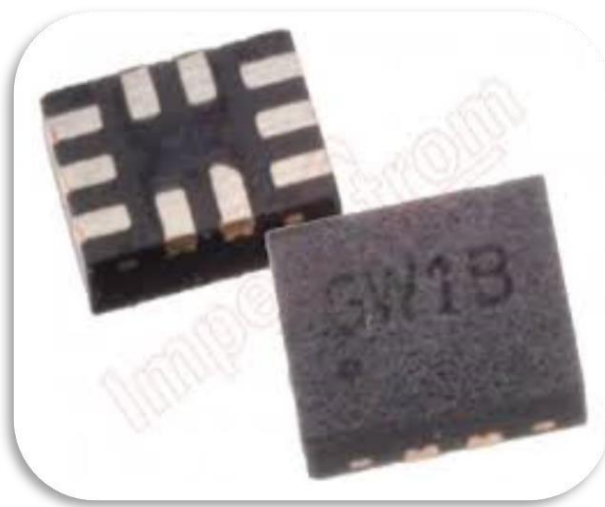

Fig8: Samsung's integrated circuits for Galaxy A5

### REFERENCES

(1) S. J. Choi, I. Lee, B. H. Jang, D. Y. Youn, W. H. Ryu, C. O. Park, I. D. Kim, Analytical Chemistry, **2013**, 85, 1792-1796
(2) Zibiao Li, Materials Science and Engineering C Functionalization of 2D TMDs for biomedical applications, 2016, 1-10.
(3) Wonbong Choi, Recent development of 2D TMDs and their applications, 2015, 3, 16419.
(4) H. Huang, X. Wang, P. Wang, G. Wu, Y. Chen, C. Meng, L. Liao, J. Wang, W. Hu, H. Shen, *RSC Adv.* **2016**, *6*, 87416.